\documentstyle[11pt,fleqn]{article}
\begin{document}

\begin{titlepage}

\begin{flushright}
Freiburg--THEP 95/11\\
June 1995
\end{flushright}
\vspace{1.5cm}

\begin{center}
\large\bf
{\LARGE\bf Two--loop heavy Higgs correction to Higgs decay
           into vector bosons}\\[1cm]
\rm
{Adrian Ghinculov}\\[.5cm]

{\em Albert--Ludwigs--Universit\"{a}t Freiburg,
           Fakult\"{a}t f\"{u}r Physik}\\
      {\em Hermann--Herder Str.3, D-79104 Freiburg, Germany}\\[1.5cm]

\end{center}
\normalsize

\begin{abstract}

The leading $m_H$ radiative correction to the Higgs decay width
into a pair of weak vector bosons is calculated at the two--loop level,
using the equivalence theorem in Landau gauge. The result
indicates the breakdown of perturbation theory if the Higgs boson
is heavier than $\sim 930$ GeV, in spite of the smallness of
the one--loop radiative correction.
\end{abstract}

\vspace{3cm}

\end{titlepage}

%############################################################################

\section{Introduction}

After the recent discovery of the top quark at the Tevatron,
the mechanism of spontaneous electroweak symmetry breaking
undoubtedly remains the most obscure part of the standard model.
The screening theorem \cite{veltman:screening,einhorn:screening},
which states that the leading contributions
in $m_H$ cancel in the radiative corrections to low energy processes,
results in the present loose limits on the Higgs mass derived
from radiative corrections.

A Higgs mass of the order of 1 TeV is definitely not excluded by
the present data. A recent analysis of the LEP results even points in the
direction of a heavy Higgs \cite{consoli}. However, other analyses
give somewhat different results \cite{hagiwara}. The uncertainty is apparently
too large to reach a conclusive estimate of the Higgs mass.

A large Higgs mass means strong interactions in
the spontaneous symmetry breaking sector, and would challenge the
theory because of the breakdown of the perturbative approach. For a
not too heavy Higgs, one may try to resum the asymptotic perturbative
series by means of Pad\'{e} approximants or by using various nonlinear
sequence transformations such as the Levin's sequence transformation.
Such techniques proved useful on simple quantum mechanical examples
\cite{vinette},
but strictly speaking there is no proof that they would converge
in the case of radiative corrections in the standard model.

At the same time, strong couplings in the spontaneous symmetry
breaking sector may result in a rich, nonperturbative spectrum
of new phenomena at the TeV energy scale. In the limit of infinite Higgs
mass, the Higgs sector reduces formally to the Lagrangian of
the nonlinear sigma model, with the role of the chiral symmetry
played by the custodial symmetry. In the low energy limit one can use
chiral perturbation theory to derive anomalous selfinteractions of
the electroweak gauge bosons \cite{longhitano}.
At the TeV energy scale, a nonperturbative
spectrum of resonances may appear by analogy with pion physics
\cite{HMveltman},
and restore partial wave unitarity in longitudinal vector boson
scattering. The BESS model was originally proposed as a possible
description of such bound states of the electroweak symmetry
breaking sector \cite{casalbuoni}.

One can estimate the validity range of perturbation
theory from unitarity violation of tree level amplitudes
\cite{dicus-mathur,lee-quigg}, and from
the size of one--loop radiative corrections
\cite{durand,marciano}. A perhaps more explicit criterion
to estimate the range beyond
which the asymptotic series stops converging at all requests
knowledge of radiative corrections at two--loop order.
The value of each term in the perturbative expansion of
an amplitude depends on the renormalization scheme
(see for instance ref. \cite{sirlin}),
and therefore the range of validity of the perturbative
approach is scheme dependent. However, for a given scheme,
at the point
where the two--loop corrections are as large as the one--loop ones,
the perturbative series shows no sign of convergence, and
there is no warranty that the tree or the one--loop level amplitudes
calculated in that scheme are reasonable approximations.

Calculating two--loop radiative corrections in the electroweak theory
is not an easy task. However, the problem becomes considerably simpler
if the external momenta can be neglected. It was shown by van der Bij
and Veltman that any two--loop Feynman diagram evaluated at vanishing
external momenta can be expressed in terms of the derivatives of the
so--called master diagram, for which they found an analytical formula
in terms of Spence functions \cite{vdBij:2loop:rho}.
This made it possible to evaluate two--loop
heavy Higgs effects on low energy observables such as the $\rho$ parameter
\cite{vdBij:2loop:rho}, the masses of the vector bosons
\cite{vdBij:2loop:masses}, and the trilinear selfcouplings of the vector bosons
\cite{vdBij:2loop:vertex}.
As expected, these effects are rather small because of the screening of the
heavy sector, and perturbation theory only breaks down for a Higgs
particle as heavy as 3---4 TeV.

To calculate two--loop corrections to
processes involving the symmetry breaking scalars, one has to evaluate
two--loop diagrams with finite external momenta. This is a more difficult
problem, and calculations of physical quantities
have not been performed until
recently, in spite of the considerable effort which was devoted
to solving massive two--loop integrals.
The reason is that in general the massive two--loop scalar diagrams
with arbitrary masses and external momenta cannot be expressed in
terms of known and easy to evaluate functions, like polylogarithms.
For instance, the two--loop selfenergy diagram with three propagators was
shown to be related to a certain generalization of the hypergeometric
series, the Lauricella function \cite{lauricella}.

A number of techniques were proposed to deal with some two--loop diagrams.
Without exhausting the list, let us merely mention a few.
Analytical results exist for certain diagrams evaluated at special values
of the masses and of the external momenta \cite{scharf,davydychev:vertex}.
Several integral representations
were proposed for some diagrams \cite{kreimer,berends,czarnecki,bauberger},
as well as Monte--Carlo integration over
the Feynman parameters \cite{fujimoto1,fujimoto2}.
Momentum expansions were worked out too
\cite{davydychev:small,davydychev:large,berends:zero}, and
conformal transformations combined with Pad\'{e} approximants or
Levin's sequence transformation were used to extend their
convergence domain \cite{fleischer}.

This paper uses a recently proposed method to deal with massive
two--loop diagrams in a systematic way \cite{2loop:method}.
It can be used, at least
in principle, to evaluate any massive two--loop scalar integral
with controllable accuracy, and is suitable for implementation in
a computer program to generate and calculate automatically the Feynman
graphs relevant for a given physical process. The same method was
used to calculate the Higgs selfenergy at two--loop level and to
extract the leading corrections to the location of the Higgs resonance
as seen in fermion scattering \cite{2loop:method}.
This gives a perturbative bound on
the Higgs mass of $\sim 1.2$ TeV. The two--loop heavy Higgs corrections
to the Higgs fermionic width give a somewhat lower perturbative bound
of $\sim 1.1$ TeV \cite{2loop:htoferm,kniehl}.

This paper presents the two--loop corrections to the $H \rightarrow Z^0 Z^0$
and $H \rightarrow W^+ W^-$ decays in the heavy Higgs limit.
More precisely, only the leading $m_H$ corrections will be
retained, which grow like $m_H^4$ at two--loop level.
As expected, these processes allow one to derive a stronger
perturbative bound on the mass of the
Higgs boson. The two--loop correction becomes as large as the one--loop
one for $m_H \sim 930$ GeV, indicating that,
at least in the OMS renormalization scheme,
the perturbation theory is not reliable anymore.

%############################################################################

\section{Lagrangian and renormalization}

Since one is only interested in radiative corrections at leading order
in the Higgs mass, the natural choice is to use the equivalence theorem
and the Landau gauge.

The equivalence theorem \cite{cornwall}---\cite{he:ET}
relates the Green functions with external longitudinal
vector bosons $V_L$ to the Green functions with the corresponding Goldstone
bosons $\phi$ replacing the vector boson legs:

\begin{equation}
           A( V_{L}^{i_{1}}, V_{L}^{i{2}}, \ldots , V_{L}^{j_{n}} ) =
  (i C)^n  A( \phi^{i_{1}}, \phi^{i_{2}}, \ldots , \phi^{i_{n}} )
        + {\cal O} (\frac{m_{W}}{\sqrt{s}})        \; \; .
\end{equation}
Such relations are a consequence of the Slavnov--Taylor identities
of the theory. The coefficient $C$ with which the amplitude has to be
multiplied for each external longitudinal vector boson replaced by a
Goldstone boson is gauge dependent and is in general not unity beyond
tree level \cite{he:ET}.

The calculation is simpler if performed
in Landau gauge. In this gauge the coefficient
$C$ in eq. 1 is one at leading order in $m_H$ \cite{he:ET},
and the diagrammatics
is considerably simplified. There are no Goldstone boson--vector boson
mixing terms, and diagrams containing fermions, vector bosons or
Fadeev--Popov ghosts need not be taken into account since they are
nonleading. The only diagrams which give leading contributions in
$m_H$ are those which contain only the symmetry breaking scalars.

The scalar sector of the standard model reads:

\begin{eqnarray}
{\cal L} & = &
\frac{1}{2} (\partial_{\mu}H_{0})(\partial^{\mu}H_{0}) +
\frac{1}{2} (\partial_{\mu}z_{0})(\partial^{\mu}z_{0}) +
            (\partial_{\mu}w_{0}^{+})(\partial^{\mu}w_{0}^{-})
                                                \nonumber \\
& & - g^{2}\frac{m_{H_{0}}^{2}}{m_{W_{0}}^{2}} \frac{1}{8} \,
[ \, w_{0}^{+} w_{0}^{-} + \frac{1}{2} z_{0}^{2} + \frac{1}{2} H_{0}^{2}
+ \frac{2 m_{W_{0}}}{g} H_{0}
+ \frac{4 \, \delta t}{g^{2} \, \frac{m_{H_{0}}^{2}}{m_{W_{0}}^{2}}}
 \, ]^{2}
      \; \; ,
\end{eqnarray}
where $H_0$, $w^{\pm}_0$ and $z_0$ denote the bare Higgs and
Goldstone fields. The gauge coupling constant $g$ can be defined
by using the muon decay as $g^{2} = 4 \sqrt{2} \, m_{W}^{2} \, G_{F}$,
with $G_{F} = 1.16637 \cdot 10^{-5} \; GeV^{-2}$, and $m_{W} = 80.22 \; GeV$.
The energy scale where the gauge coupling constant
is defined is actually irrelevant within this context because $g$
is not renormalized at leading order in $m_H$.
$\delta t$ is the tadpole counterterm which will be fixed by the condition that
the vacuum expectation value of the  Higgs field, $v$ ,
should not receive quantum corrections, and is related to the selfenergy
of the Goldstone bosons at zero momentum transfer \cite{taylor}.

To renormalize the theory, one splits the bare Lagrangian of eq. 2
into the renormalized Lagrangian and counterterms:

\begin{eqnarray}
H_{0} & = & Z_{H}^{1/2} H      \nonumber \\
z_{0} & = & Z_{G}^{1/2} z      \nonumber \\
w_{0} & = & Z_{G}^{1/2} w      \nonumber \\
m_{H_{0}}^{2} & = & m_{H}^{2} - \delta m_{H}^{2}     \nonumber \\
m_{W_{0}}^{2} & = & m_{W}^{2} - \delta m_{W}^{2}
      \; \; ,
\end{eqnarray}

We use an on--shell renormalization scheme with field renormalization.
In this renormalization scheme, the counterterms can be determined from
the following renormalization conditions:

\begin{eqnarray}
 & & \hat{\Sigma}_{HH} (k^2=m_H^2) + i \, \delta m_H^2 - i \, \delta t
+ i \, \delta m_H^2 \, \delta Z_H - i \, \delta t \, \delta Z_H
 =  0  \nonumber \\
 & & \frac{\partial}{\partial k^2} \hat{\Sigma}_{HH} (k^2=m_H^2)
 + i \, \delta Z_H  =  0      \nonumber \\
 & & \hat{\Sigma}_{w^+ w^-} (k^2=0)
- i \, \delta t - i \, \delta t \, \delta Z_G  =  0      \nonumber \\
 & & \frac{\partial}{\partial k^2} \hat{\Sigma}_{w^+ w^-} (k^2=0)
+ i \, \delta Z_G   =  0     \nonumber \\
 & & \hat{\Sigma}_{W^+ W^-} (k^2=0) + i \, \delta m_W^2   =  0
      \; \; ,
\end{eqnarray}
where $\hat{\Sigma}$ contain the loop and loop--counterterm selfenergy
diagrams,
but not the pure counterterm diagrams.
$\hat{\Sigma}_{W^+ W^-} (k^2)$ is the coefficient of the $-g_{\mu \nu}$
piece of the vector boson selfenergy. Because of the Ward identity
$\delta m_{W}^{2} = - m_{W}^{2} \delta Z_{G}$ , it's not really necessary
to consider the gauge sector to calculate the mass counterterm of
the vector bosons, as done in the last line of eqns. 4, but this was done
nevertheless because it provides a useful check on the calculation.
Another check which was done is to compute the Higgs tadpole diagrams
and to check that they cancel upon the tadpole counterterm
derived from the Goldstone boson selfenergy at zero momentum transfer.

At one--loop level, one needs to evaluate the H--H, w--w and W--W selfenergies
whose topologies are shown in fig. 1, to find the following counterterms at
order $g^{2} \; \frac{m_{H}^{2}}{m_{W}^{2}}$:

\begin{eqnarray}
\delta t^{(1-loop)} & = &  g^{2} \frac{m_{H}^{2}}{m_{W}^{2}} \,
                           (\frac{m_{H}^{2}}{4 \pi \mu^{2}})^{\epsilon /2} \,
                           \frac{m_{H}^{2}}{16 \pi^{2}} \, \{
- {{3}\over {4\,\epsilon }} + {{3}\over 8}  -
  {{3\, \gamma }\over 8}
     \nonumber \\
& &  + \epsilon \,\left( {{-3}\over {16}} +
     {{3\, \gamma }\over {16}} -
     {{3\,{{ \gamma }^2}}\over {32}} -
     {{{{\pi }^2}}\over {64}} \right)
  \}
     \nonumber \\
\delta m_{H}^{2\,(1-loop)} & = & g^{2} \frac{m_{H}^{2}}{m_{W}^{2}} \,
                           (\frac{m_{H}^{2}}{4 \pi \mu^{2}})^{\epsilon /2} \,
                           \frac{m_{H}^{2}}{16 \pi^{2}} \, \{ \,
{{3}\over {\epsilon }} - 3 +
  {{3\, \gamma }\over 2} +
  {{{3\,\sqrt{3}}\,\pi }\over 8}
     \nonumber \\
& &  + \epsilon \,[ 3 -
     {{3\, \gamma }\over 2} +
     {{3\,{{ \gamma }^2}}\over 8} -
     {{{3\,\sqrt{3}}\,\pi }\over 8} +
     {{{3\,\sqrt{3}}\, \gamma \,
         \pi }\over {16}}
     \nonumber \\
& &  - {{{{\pi }^2}}\over {16}} -
     {{{3\,\sqrt{3}}\,
         {\it Cl}({{\pi }\over 3})}\over 4} +
     {{{3\,\sqrt{3}}\,\pi \,\log (3)}\over
       {16}}  \, ]
  \}
     \nonumber \\
\delta m_{W}^{2\,(1-loop)} & = & g^{2} \frac{m_{H}^{2}}{m_{W}^{2}} \,
                           (\frac{m_{H}^{2}}{4 \pi \mu^{2}})^{\epsilon /2} \,
                           \frac{m_{W}^{2}}{16 \pi^{2}} \, [ \,
{{{ 1 }}\over 8} +
  \epsilon \,\left( - {{3}\over {32}} +
     {{ \gamma }\over {16}} \right)
 \, ]
     \nonumber \\
\delta Z_{H}^{(1-loop)} & = &  g^{2} \frac{m_{H}^{2}}{m_{W}^{2}} \,
                           (\frac{m_{H}^{2}}{4 \pi \mu^{2}})^{\epsilon /2} \,
                           \frac{1}{16 \pi^{2}} \, \{ \,
{3\over 2} -
    {{\pi \,{\sqrt{3}}\over 4}} + \epsilon \,
   [ \, -{3\over 2} + {{3\, \gamma }\over 4}
        \nonumber \\
& & + {{{3\,\sqrt{3}}\,\pi }\over {16}} -
    {{{\sqrt{3}}\, \gamma \,\pi }\over 8} +
     {{{\sqrt{3}}\,{\it Cl}({{\pi }\over 3})}\over 2} -
     {{{\sqrt{3}}\,\pi \,\log (3)}\over 8} \, ]
  \}
     \nonumber \\
\delta Z_{G}^{(1-loop)} & = &  g^{2} \frac{m_{H}^{2}}{m_{W}^{2}} \,
                           (\frac{m_{H}^{2}}{4 \pi \mu^{2}})^{\epsilon /2} \,
                           \frac{1}{16 \pi^{2}} \, [ \,
-{1\over 8} + \epsilon \,
   \left( {3\over {32}} -
     {{ \gamma }\over {16}} \right)
 \, ]
      \; \; .
\end{eqnarray}

In these expressions the space--time dimension is $n = 4 + \epsilon$, and
$Cl$ is the Clausen function, $Cl(x) = \sum_{n=1}^{\infty} \sin{(nx)} /n^2$.
Numerically, $Cl(\frac{\pi}{3}) = 1.0149416064\dots$ .
The one--loop counterterms are needed at order $\epsilon$ because
these terms combine with the poles of one--loop scalar integrals
to give finite contributions at two--loop order.

By evaluating the two-loop H--H, w--w, and W--W selfenergies, one finds
the following ${\cal O}((g^{2} \frac{m_{H}^{2}}{m_{W}^{2}})^2)$ counterterms
\cite{2loop:method,2loop:htoferm}:

\begin{eqnarray}
\delta t^{(2-loop)} & = &  (g^{2} \frac{m_{H}^{2}}{m_{W}^{2}})^{2} \,
                           (\frac{m_{H}^{2}}{4 \pi \mu^{2}})^{\epsilon} \,
                           \frac{m_{H}^{2}}{(16 \pi^{2})^{2}} \, [ \,
 {{45}\over {16\,{{\epsilon }^2}}} +
 \frac{1}{\epsilon}    ( -{{33}\over 8}
     \nonumber \\
& &  + {{45\, \gamma  }\over
        {16}} + {{{9\,\sqrt{3}}\,\pi }\over {16}} ) +
  {{609}\over {128}} - {{33\, \gamma  }\over 8} +
  {{45\,{{ \gamma  }^2}}\over {32}} -
  {{45\,\sqrt{3}\,\pi }\over {64}}
     \nonumber \\
& &    +  {{{9\,\sqrt{3}}\, \gamma  \,\pi }\over
    {16}} - {{3\,{{\pi }^2}}\over {32}}
     - {{21\,\sqrt{3}\,
      {\it Cl}({{\pi }\over 3})}\over {32}} +
  {{{9\,\sqrt{3}}\,\pi \,\log (3)}\over {32}}
  \, ]
   \nonumber  \\
\delta m_{W}^{2 \, (2-loop)} & = & - (g^{2} \frac{m_{H}^{2}}{m_{W}^{2}})^{2} \,
                           (\frac{m_{H}^{2}}{4 \pi \mu^{2}})^{\epsilon} \,
                           \frac{m_{W}^{2}}{(16 \pi^{2})^{2}} \, [ \,
 \frac{3}{32} \frac{1}{\epsilon}
 - \frac{1}{128} + \frac{3}{32} \gamma -
     \nonumber \\
  & & - \frac{\pi^2}{192}
 + \frac{3 \sqrt{3} \pi}{64} - \frac{3 \sqrt{3}}{16} Cl(\frac{\pi}{3})
  \, ]
     \nonumber \\
\delta m_{H}^{2 \, (2-loop)} & = &  Re \{ \,
     (g^{2} \frac{m_{H}^{2}}{m_{W}^{2}})^{2} \,
     (\frac{m_{H}^{2}}{4 \pi \mu^{2}})^{\epsilon} \,
     \frac{m_{H}^{2}}{(16 \pi^{2})^{2}} \, [ \,
 - {{9}\over {{{\epsilon }^2}}} \nonumber \\
  & & + \frac{3}{32\,\epsilon }  ( 169 - 96\,\gamma -  24\,\sqrt{3}\,\pi  )
     \nonumber \\
& & \; \; \; \; \; \; - (4.785031 \pm 4.2 \cdot 10^{-5})
                 - i \, (0.412438 \pm 1.6 \cdot 10^{-5})
   \, ] \, \}
     \nonumber \\
\delta Z_{G}^{(2-loop)} & = &   (g^{2} \frac{m_{H}^{2}}{m_{W}^{2}})^{2} \,
                           (\frac{m_{H}^{2}}{4 \pi \mu^{2}})^{\epsilon} \,
                           \frac{1}{(16 \pi^{2})^{2}} \, [ \,
 \frac{3}{32} \frac{1}{\epsilon}
 - \frac{1}{128} + \frac{3}{32} \gamma -
     \nonumber \\
  & & - \frac{\pi^2}{192}
 + \frac{3 \sqrt{3} \pi}{64} - \frac{3 \sqrt{3}}{16} Cl(\frac{\pi}{3})
  \, ]
     \nonumber \\
\delta Z_{H}^{(2-loop)} & = & Re \{ \,
                           (g^{2} \frac{m_{H}^{2}}{m_{W}^{2}})^{2} \,
                           (\frac{m_{H}^{2}}{4 \pi \mu^{2}})^{\epsilon} \,
                           \frac{1}{(16 \pi^{2})^{2}}
 \, [ \, \frac{3}{32} \frac{1}{\epsilon} -
     \nonumber \\
 & &          - (0.62296 \pm 2.5 \cdot 10^{-4})
         - i \, (1.00233 \pm 2.5 \cdot 10^{-4})
 \, ] \, \} \,
      \; \; ,
\end{eqnarray}

These counterterms agree with an independent calculation of the Higgs and
Goldstone selfenergies by P.N. Maher, L. Durand, and K. Riesselmann
\cite{riesselmann}.

%############################################################################

\section{The calculation}

The partial decay widths of the Higgs boson into a pair of
vector bosons are given at tree level by:

\begin{eqnarray}
\Gamma (H \rightarrow W^+ W^-) & = &
 \frac{g^2}{64 \pi} \frac{m_H^3}{m_W^2}
 \left[ 1 - 4 \frac{m_W^2}{m_H^2} \right]^{1/2}  \times
  \nonumber \\  & &
 \left[ 1 - 4 \frac{m_W^2}{m_H^2} + 12 \frac{m_W^4}{m_H^4} \right]
    \nonumber \\
\Gamma (H \rightarrow Z^0 Z^0) & = &
 \frac{g^2}{128 \pi} \frac{m_H^3}{m_W^2}
 \left[ 1 - 4 \frac{m_Z^2}{m_H^2} \right]^{1/2}  \times
   \nonumber \\  & &
 \left[ 1 - 4 \frac{m_Z^2}{m_H^2} + 12 \frac{m_Z^4}{m_H^4} \right]
\end{eqnarray}

The one--loop radiative corrections at leading order in $m_H$
can be calculated from the diagrams
shown in fig. 3. The resulting $H w^+ w^-$ coupling,
including the order $g^2 \, m_H^2 / m_W^2$ corrections, is:

\begin{eqnarray}
 &  & - i \frac{g}{2} \frac{m_H^2}{m_W}
 \left\{ 1 + \frac{g^2}{16 \, \pi^2} \frac{m_H^2}{m_W^2}
     \left[ \frac{19}{16} + \frac{5 \, \pi^2}{48}
          - \frac{3 \, \sqrt{3} \, \pi}{8} \right. \right. \nonumber \\
 & &  \left. \left.
     + i \, \pi \left( \frac{\log{2}}{4} - \frac{5}{8} \right)
      \right]  \right\} \; \; \; ,
\end{eqnarray}
%\begin{equation}
% - i \frac{g}{2} \frac{m_H^2}{m_W}
% \left\{ 1 + \frac{g^2}{16 \, \pi^2} \frac{m_H^2}{m_W^2}
%     \left[ \frac{19}{16} + \frac{5 \, \pi^2}{48}
%          - \frac{3 \, \sqrt{3} \, \pi}{8}
%     + i \, \pi \left( \frac{\log{2}}{4} - \frac{5}{8} \right)
%      \right]  \right\} \; \; \; ,
%\end{equation}
and the partial widths of eq. 7 correspondingly get a correction factor

\begin{equation}
        1 + \frac{g^2}{16 \, \pi^2} \frac{m_H^2}{m_W^2}
     \left( \frac{19}{8} + \frac{5 \, \pi^2}{24}
          - \frac{3 \, \sqrt{3} \, \pi}{4}
    \right)  \; \; \; .
\end{equation}

The real part of the correction of eq. 8 and the corresponding
correction to the widths of eq. 9 agree with the results of ref.
\cite{marciano}.
Note that at two--loop level one needs the imaginary part of the
one--loop radiative correction of eq. 8 as well because it gives
a correction to the widths of order $(g^2 \, m_H^2 / m_W^2)^2$.

One problem related to the use of the Landau gauge is the presence of
massless particles in the theory. Some Feynman diagrams may display
mass singularities, and there is the problem of the arbitrariness of
integrals of the type $\int d^n p \; (1/p^4)$ in the framework of dimensional
regularization. Such problems do not appear in this calculation at one--loop
level, since the only place where one encounters
a $\int d^n p \; (1/p^4)$ type
integral is the $k_\mu k_\nu$ piece of the vector boson propagator.
They are however present at two--loop level, and one needs a regularization
procedure to deal with these problems in a consistent way.
The approach adopted in this paper
is to work in a nearly--Landau gauge, that is, to keep a small gauge parameter
$\xi$ during the calculation, and to let $\xi \rightarrow 0$ in the final
results. This amounts to giving the Goldstone bosons a small mass
$\sqrt{\xi} \; m_W$.
This procedure is consistent with discarding the diagrams containing
fermions, gauge bosons and Fadeev--Popov ghosts on the internal lines,
since they do not give rise to contributions of
${\cal O}((g^2 \; m_H^2/m_W^2)^2)$ in the limit $\xi \rightarrow 0$.
In this way, an additional check on the calculation is present,
namely the cancellation of the poles and logarithms of the gauge
parameter in the final result. This cancellation involves both the
analytical and the numerical parts of the calculation.
Another possibility, advocated in ref. \cite{riesselmann}, is to calculate
the singular diagrams with vanishing Goldstone boson masses but
at off--shell momentum, and to check that their sum remains finite
when the external momentum is put on--shell.

Let us now turn to the actual two--loop calculation. The main task
is to calculate the two--loop proper diagrams corresponding
to the $H w^+ w^-$ vertex.
The topologies which are involved are shown in fig. 4.

Considering the large number of Feynman diagrams which need to be
calculated and the lengthy expressions which are involved at
intermediary steps, the whole calculation was done by computer.
The algorithm is essentially the same which was used to derive
the two--loop counterterms given in the previous section.

As a first step, a computer program generates all relevant
Feynman diagrams. This was done by giving by hand the possible
topologies of the proper vertex diagrams, which are shown in fig. 4.
For each given topology, the program then substitutes
for each internal line all possible propagators, that is, $H$--$H$,
$z$--$z$, $w^+$--$w^-$, and $w^-$--$w^+$. This way many diagrams are generated
which contain nonexistent vertices. The program then compares the
vertices of the diagrams generated with a complete list of the vertices of
the theory, discards the spurious diagrams, and substitutes the actual
expressions of the vertices in the correct diagrams. The combinatorial
factors are automatically correct, provided one divides each topology by
its symmetry factor. If a certain topology has $m$ sets of equivalent
lines or nodes which contain $n_1, n_2, \dots , n_m$ elements, then
its symmetry factor is $n_1 ! \; n_2 ! \; \dots \; n_m !$.

In the next step some algebra is necessary to bring the resulting
Feynman diagrams into a standard form \cite{2loop:method}. After doing
all the partial fractioning which is possible, the program decides which
diagrams can be calculated analytically and which ones need numerical
integration. The two--loop diagrams evaluated at vanishing external momentum
can be expressed analytically in terms of Spence functions,
and in this case their mass expansion is needed in order to retain
only the ${\cal O}((g^2 \; m_H^2/m_W^2)^2)$ terms in the final expression.
In the case of the two--loop diagrams evaluated at finite external momenta,
the propagators with the same loop momentum are combined by means of Feynman
parameters, and the resulting scalar integrals are calculated in terms of
two basic functions, $g$ and $f$, whose definitions are given in the Appendix.

Some care is needed when introducing Feynman parameters in triangular
diagrams. It is useful to parametrize in the same way the diagrams
with a similar structure in order to avoid the unnecessary proliferation
of expressions resulting from introducing Feynman parameters.
There are two types of triangular diagrams which appear in this
calculation, and they can be parametrized in the following way:

\begin{eqnarray}
\lefteqn{\frac{1}{[(p-k_1)^2 - M^2]^{\alpha_1}
         [(p+k_2)^2 - M^2]^{\alpha_2}
         [p^2 - m^2]^{\alpha_3} } \, = \,
    \frac{\Gamma(\alpha_1+\alpha_2+\alpha_3)}{\Gamma(\alpha_1)
                               \Gamma(\alpha_2)\Gamma(\alpha_3)}
    } & & \nonumber \\
 & &  \times \int_0^1 dx \, \int_0^1 dy \,
 \frac{x (xy)^{\alpha_1-1} [x (1-y)]^{\alpha_2-1} (1-x)^{\alpha_3-1}}{
            [(p+\tilde{k}_1)^2-m_1^2+i\eta]^{\alpha_1+\alpha_2+\alpha_3}}  \\
\lefteqn{\frac{1}{[(p-k_1)^2 - m^2]^{\alpha_1}
         [p^2 - M^2]^{\alpha_2}
         [(p+k_2)^2 - m^2]^{\alpha_3} } \, = \,
    \frac{\Gamma(\alpha_1+\alpha_2+\alpha_3)}{\Gamma(\alpha_1)
                               \Gamma(\alpha_2)\Gamma(\alpha_3)}
    } & & \nonumber \\
 & &  \times \int_0^1 dx \, \int_0^1 dy   \,
 \frac{x (xy)^{\alpha_1-1} [x (1-y)]^{\alpha_2-1} (1-x)^{\alpha_3-1}}{
            [(p+\tilde{k}_2)^2-m_2^2+i\eta]^{\alpha_1+\alpha_2+\alpha_3}}  \;
\; ,
\end{eqnarray}
where $k_1$ and $k_2$ are the momenta of the external Goldstone bosons,
$M \equiv m_H$ and $m \equiv \sqrt{\xi} \, m_W$ are the masses of the Higgs
and Goldstone bosons, and

\begin{eqnarray}
\tilde{k}_1 & = & - x y \, k_1 + x (1-y) \, k_2  \nonumber \\
\tilde{k}_2 & = & - x y \, k_1 + (1-x) \, k_2  \nonumber \\
m_1^2       & = &  x[1-xy(1-y)] \, M^2 + (1-x) \, m^2  \nonumber \\
m_2^2       & = &  x[1-y(2-x)]  \, M^2 + [1-x(1-y)] \, m^2   \; \; .
\end{eqnarray}

The diagrams containing the second structure have at least one
two--particle cut, and correspondingly an imaginary part.
After introducing Feynman parameters, this translates into the presence
of poles which one has to avoid by a suitable choice of the
integration path. With respect to the
numerical integration over the Feynman parameters $x$ and $y$,
the parametrization given in eq. 11 has the advantage
that the solution of the equation $m_2 = 0$ has a simple structure
in the integration domain $(0,1) \times (0,1)$, and therefore it is easy to
find an integration path to avoid this singularity.

Also related to this parametrization is the problem of
using the Landau gauge to calculate the diagrams of fig. 5.
Such diagrams taken separately are not well defined for physical momenta
because of an endpoint singularity,
and it is only their sum which is finite. More precisely, these
diagrams have two two--particle cuts, and after introducing
two Feynman parameters, one obtains an expression of the type in
eq. 11. The singularities of the integrand in the $xy$ plane
lie on a curve given by $m_2 = 0$, which intersects the boundaries
of the integration domain $(0,1) \times (0,1)$. These singularities can
be avoided easily by deforming the integration paths in a convenient way
inside the square $(0,1) \times (0,1)$, but not on its frontiers, and
therefore the integral is logarithmically divergent. One way to regularize
this divergency is to calculate these diagrams slightly off--shell,
for instance by keeping a small but finite $\eta$ in eq. 11, and to
take the limit $\eta \rightarrow 0$ in the sum of these three diagrams.
The problem is that also the Landau gauge limit must be taken at the
end of the calculation, and this must be handled with care because
these two limits do not commute. This is due to
the presence of pieces which are
nonleading in $m_H$ and ought to vanish in Landau gauge, but which
display also the endpoint logarithmic singularity.
Therefore, the correct way to take the limit is first
to set $\xi \rightarrow 0$,
and then to go on--shell.

A simple way to get rid of this problem
is to introduce explicitly in the diagram of fig. 5 c) the terms
$\xi \, m_W^2 \, \delta Z_G \, w^+ w^-$ and
$1/2 \,\xi \, m_W^2 \, \delta Z_G \, z z$,
which in fact exist in the Lagrangian
of eq. 2 when one is not exactly in Landau gauge, but which only
generate contributions which vanish when $\xi \rightarrow 0$.
Taking these terms into account in the diagram 5 c) ensures the cancellation of
the nonleading contributions which have the endpoint singularity, after
which the order in which one takes the limits $\xi \rightarrow 0$
and $\eta \rightarrow 0$ becomes irrelevant.
Since the endpoint divergency is only logarithmic, no large numerical
cancellations appear. Moreover, the cancellations among the diagrams
already occur at
the level of the integrands, before the numerical integration
is carried out, because the diagrams were parametrized in the same way.
The result is thus numerically stable.

Some checks on the algebraic part were also included in the program.
Where possible, the analytical cancellation of the poles and
logarithms of the gauge parameter was checked. Some subsets of
diagrams must be free of ultraviolet divergencies, in agreement
with Bogoliubov's proof of renormalizability. This is the case with
the combinations $c+d+r$, $k+l+s$, and $g+h+i+j+o$, where the notations
of topologies are defined according to fig. 4.
Note that in order to check this, one needs
first to set the one--loop field renormalization counterterms to zero.

The algorithm described was encoded in a FORM \cite{vermaseren} program.
It takes approximately seven hours to generate all Feynman diagrams, and
to perform the necessary algebra and the checks on a NeXT computer.
In the end, one obtains an analytical part and a number of numerical
integrals over Feynman parameters of $g$ functions, which were
encoded in FORTRAN. Their evaluation took approximately 10 hours on
an Apollo 9000/720 workstation.

In the end, one obtains the following result for the $H w^+ w^-$
coupling, including the order $(g^2 \; m_H^2/m_W^2)^2$ radiative corrections:

\begin{eqnarray}
 &  & - i \frac{g}{2} \frac{m_H^2}{m_W}
 \left\{ 1 + \frac{g^2}{16 \, \pi^2} \frac{m_H^2}{m_W^2}
     \left[ \frac{19}{16} + \frac{5 \, \pi^2}{48}
          - \frac{3 \, \sqrt{3} \, \pi}{8} \right.
% \right. \nonumber \\
% & &   \left.
     + i \, \pi \left( \frac{\log{2}}{4} - \frac{5}{8} \right)
    \right]  \nonumber \\
 & & + \left( \frac{g^2}{16 \, \pi^2} \frac{m_H^2}{m_W^2} \right)^2
      \left.   \left[ - (.53673 \pm 4.1 \cdot 10^{-4} ) \right. \right.
            \nonumber \\
 & &  \left. \left.   - i \, (.32811 \pm 3.1 \cdot 10^{-4} ) \, \right] \,
     \right\} \; \; \; .
\end{eqnarray}

Correspondingly, the partial widths of eq. 7 get
the following correction factor:

\begin{eqnarray}
 & &       1 + \frac{g^2}{16 \, \pi^2} \frac{m_H^2}{m_W^2}
     \left( \frac{19}{8} + \frac{5 \, \pi^2}{24}
          - \frac{3 \, \sqrt{3} \, \pi}{4}
    \right)  \nonumber \\
 & &   \; \; \; \; \; \; \; \; \; \; \; \; \; \; \; \; \; \; \; \; \;
       + \left(\frac{g^2}{16 \, \pi^2} \frac{m_H^2}{m_W^2} \right)^2
     \left( \, .97103 \pm 8.2 \cdot 10^{-4} \, \right)
     \; \; \; .
\end{eqnarray}

This correction factor is shown in fig. 6 as a function of $m_H$.
The two--loop correction has the same sign as the one--loop correction,
and for $m_H \approx 930$ GeV it becomes as large as the latter.
This is an indication for the validity range of perturbation theory.
This is a rather
surprising result, taking into account that the one--loop radiative
correction is quite small for such a Higgs mass,
at the level of $\sim 13\%$.

\begin{table}
\begin{tabular}{||c||c|c|c|c||}                       \hline\hline
 $m_{H}$   & $\Gamma(H \rightarrow W^+W^-)$
           & $\Gamma(H \rightarrow Z^+Z^-)$
	   & $\Gamma(H \rightarrow t\bar{t})$
	   & total   \\
 $[$GeV$]$ &  $[$GeV$]$ & $[$GeV$]$ & $[$GeV$]$  & $[$GeV$]$  \\ \hline\hline
   400     &  16.97     & 7.920     & 2.146      & 27.04      \\ \hline
   500     &  36.82     & 17.60     & 10.89      & 65.32      \\ \hline
   600     &  68.43     & 33.16     & 20.15      & 121.7      \\ \hline
   700     &  115.7     & 56.55     & 29.09      & 201.4      \\ \hline
   800     &  184.0     & 90.38     & 37.59      & 311.9      \\ \hline
   900     &  280.5     & 138.3     & 45.58      & 464.4      \\ \hline
   1000    &  415.4     & 205.4     & 52.92      & 673.7      \\ \hline\hline
\end{tabular}
\caption{The main decay channels of a heavy Higgs boson
at two--loop order ($m_t = 180$ GeV).}
\end{table}

The partial decay widths corresponding to the main
decay channels of the Higgs boson, including the two--loop
${\cal O}((g^2 \; m_H^2/m_W^2)^2)$ radiative corrections, are
given in table 1. The $t\bar{t}$ channel, which
was calculated with similar methods in ref. \cite{2loop:htoferm},
is also given. It should be noted that by multiplying the tree level decay
rates in eqns. 7 by the correction factor of eq. 14 some subleading terms
are also generated. They start with ${\cal O}(g^2)$ terms in the one--loop
correction, and with ${\cal O}(g^4 \; m_H^2/m_W^2)$ terms at two--loop.
Such contributions are of course incomplete, but they were not explicitly
subtracted from the numerical results given in table 1 because they are
formally of the same order as the theoretical uncertainty due to the
full subleading
contributions they are part of, and also numerically
negligible. A similar discussion holds for the $t\bar{t}$ channel as well.

%############################################################################

\section{Conclusions}

The decays $H \rightarrow W^+ W^-$ and $H \rightarrow Z^0 Z^0$ were
calculated at two--loop level in the limit of large Higgs mass.
The calculation was performed in Landau gauge and by using the
equivalence theorem, in order to obtain the leading $m_H$ contributions.

The two--loop radiative corrections have the same sign as the
one--loop ones, and thus result in an enhancement of the Higgs width.
The two--loop corrections become as large as the one--loop ones for
$m_H = 930$ GeV. For this value of the Higgs mass,
the sum of one--loop and two--loop
radiative corrections is as large as 26\% of the tree level widths.
This is indicative for the point beyond which the perturbative series
stops converging at all in this renormalization scheme,
and calculations performed by means of Feynman diagrams
become unreliable.

This result is rather surprising, considering that
for $m_H = 930$ GeV the one--loop
corrections are quite small, at 13\% level. They only become substantial
for a Higgs as heavy as 1.3 TeV \cite{marciano}.
Considering that most of the existing calculations
in the electroweak theory were done in the OMS renormalization scheme,
this raises the question of the validity
range of the calculations of other processes involving the spontaneous
symmetry breaking sector, such as the $W W \rightarrow W W$ scattering which
is of interest in view of searches for the Higgs boson at future
hadron colliders.

%############################################################################

\vspace{.5cm}

{\bf Acknowledgement}

The author gratefully acknowledges interesting discussions
with prof. Jochum van der Bij.

%############################################################################
%############################################################################
%############################################################################

\appendix
\section*{Appendix}

Here we give some details related to the use of the techniques of ref.
\cite{2loop:method} to evaluate the scalar integrals needed
for this calculation.

The following two basic functions
were introduced in ref. \cite{2loop:method}:

\begin{eqnarray}
g(m_{1},m_{2},m_{3};k^2) & = &   \int_{0}^{1}\,dx\,
     [ \, Sp(\frac{1}{1-y_{1}}) + Sp(\frac{1}{1-y_{2}})
                                                \nonumber \\
& & + y_{1}\log \frac{y_{1}}{y_{1}-1} +
      y_{2}\log \frac{y_{2}}{y_{2}-1} \, ]
      \; \; ,    \\
f(m_{1},m_{2},m_{3};k^2) & = &  \int_{0}^{1}\,dx\,
     [ \, \frac{1-\mu^{2}}{2 \kappa^{2}}
                                                \nonumber \\
& & - \frac{1}{2} \, y_{1}^{2} \, \log \frac{y_{1}}{y_{1}-1}
    - \frac{1}{2} \, y_{2}^{2} \, \log \frac{y_{2}}{y_{2}-1} \,  ]
      \; \; ,
\end{eqnarray}
where

\begin{eqnarray}
y_{1,2} & = & \frac{1 + \kappa^{2} - \mu^{2}
                    \pm \sqrt{\Delta}}{2 \kappa^{2}}  \nonumber \\
\Delta  & = & (1 + \kappa^{2} - \mu^{2})^{2}
          + 4 \kappa^{2} \mu^{2} - 4 i \kappa^{2} \eta
      \; \; ,
\end{eqnarray}
and

\begin{eqnarray}
   \mu^{2}  & = &  \frac{a x + b (1-x)}{x (1-x)}   \nonumber \\
         a  & = &  \frac{m_{2}^{2}}{m_{1}^{2}} \, , \; \; \; \;
         b \; = \; \frac{m_{3}^{2}}{m_{1}^{2}} \, , \; \; \; \;
\kappa^{2} \; = \; \frac{    k^{2}}{m_{1}^{2}}
      \; \; .
\end{eqnarray}

No attempt was done to further integrate these functions analytically,
since this cannot be done in terms of known and easy to calculate functions,
such as polylogarithms. It is presumably possible to relate them
to the Lauricella functions \cite{lauricella}, but
it is not clear that this would lead to an efficient way to calculate
them. Instead, a FORTRAN routine was written to integrate numerically
 the $g$ and $f$
functions, as well as the necessary derivatives of $g$, to the desired
accuracy. This can be done easily by using an adaptative deterministic
integration algorithm.

Some tricks were used to perform the integration
in an efficient way. First, one extracts the singularities at the ends
of the integration path, which are of logarithmic type, by a convenient
change of variables, such as $t = \sqrt{x}$. Then, the program chooses
the appropriate integration path in the complex plane of the integration
variable $x$, in order to avoid the eventual singularities of the integrand.
The aim is twofold. First, on such a path the integrand has a small
variance and is smooth, so the integral can be calculated to high accuracy
by using a small number of points. Second, by carrying out the integration
on such a path, one avoids automatically the numerical instabilities
due to large cancellations which occur in the computation of the integrand
near its branching points. To compute the suitable integration
path, one needs to know that the integrands of the $g$ and $f$ functions
given in eq. 15 and 16
have two branching points in the plane of the complex Feynman parameter $x$.
They lie on the real axis when the functions are calculated
above the physical threshold
 $-k^2>(m_1+m_2+m_3)^2$, and their location is given by:

\begin{eqnarray}
x_{1,2} & = & \frac{1}{2 \mu_{1}^{2}}
             \, [ \, -a + b + \mu_{1}^{2}
   \pm \sqrt{(a-b-\mu_{1}^{2})^{2} - 4 b \mu_{1}^{2}} \, ]  \nonumber \\
\mu_{1,2}^{2}  & = & 1 - \kappa^{2} \mp 2 \sqrt{- \kappa^{2}}
      \; \; .
\end{eqnarray}

The functions $g$ and $f$ are the finite parts of the following
scalar integrals:

\begin{eqnarray}
\lefteqn{{\cal G}(m_{1},m_{2},m_{3};k^2) \, \equiv}  \nonumber \\
& & \int d^{n}p\,d^{n}q\,
       \frac{1}{
             (p^{2}+m_{1}^{2})^{2} \,
             [(q+k)^{2}+m_{2}^{2}] \,
             [(p+q)^{2}+m_{3}^{2}]
	    } \,  =  \nonumber \\
& &   \pi^{4} \{ \,   \frac{2}{\epsilon^{2}}
    + \frac{1}{\epsilon} [- 1 + 2 \gamma + 2 \log (\pi \, m_{1}^{2}) ]
    + \frac{1}{4} + \frac{\pi^{2}}{12}
                                                \nonumber \\
& & + \frac{1}{4} [- 1 + 2 \gamma + 2 \log (\pi \, m_{1}^{2}) ]^{2}
    - 1 + g(m_{1},m_{2},m_{3};k^2)
          \,    \}  \; \; ,     \\
\lefteqn{{\cal F}(m_{1}, m_{2}, m_{3} ;k^2) \, \equiv}  \nonumber \\
& & - \int d^{n}p\,d^{n}q\,
       \frac{(p+q).k}{
             (p^{2}+m_{1}^{2})
             [(q+k)^{2}+m_{2}^{2}]
             (r^{2}+m_{3}^{2})^{2}
	    } \, =  \nonumber \\
& & k^{2} \pi^{4} \{ \, - \frac{1}{2 \epsilon}
                     + \frac{9}{8}
    - \frac{1}{2} [ \gamma + \log (\pi m_{1}^{2}) ]
    + f(m_{1},m_{2},m_{3};k^2)
          \,    \}  \; \; ,
\end{eqnarray}

Let us introduce the following notation:

\begin{eqnarray}
\lefteqn{G(m_{1},\alpha_{1};
           m_{2},\alpha_{2};
           m_{3},\alpha_{3};k^2)  \, =}  \nonumber \\
& &    \int d^{n}p\,d^{n}q\,
       \frac{1}{
             (p^{2}+m_{1}^{2})^{\alpha_{1}} \,
             (q^{2}+m_{2}^{2})^{\alpha_{2}} \,
             [(r+k)^{2}+m_{3}^{2}]^{\alpha_{3}}
	    }
    \; \; .
\end{eqnarray}

All $G$ scalar integrals can be obtained from ${\cal G}$ and ${\cal F}$
with the help of the following relations:

\begin{eqnarray}
\lefteqn{G(m_{1},\alpha_{1}+1;
           m_{2},\alpha_{2};
           m_{3},\alpha_{3};k^2) \, =}  \nonumber \\
& &  - \frac{1}{\alpha_{1}} \frac{\partial}{\partial m_{1}^{2}}
        G(m_{1},\alpha_{1};
          m_{2},\alpha_{2};
          m_{3},\alpha_{3};k^2)
    \; \; ,   \nonumber \\
\lefteqn{G(m_{1},1;m_{2},1;m_{3},1;k^2) \, =}  \nonumber \\
& &  \frac{1}{3-n}
     \{   m_{1}^{2} \, {\cal G}(m_{1}, m_{2}, m_{3} ;k^2)
        + m_{2}^{2} \, {\cal G}(m_{2}, m_{1}, m_{3} ;k^2)
	                  \nonumber \\
& &  \; \; \; \;   + m_{3}^{2} \, {\cal G}(m_{3}, m_{1}, m_{2} ;k^2)
        +           {\cal F}(m_{1}, m_{2}, m_{3} ;k^2)
     \}
      \; \; ,
\end{eqnarray}

We further notice that any two--loop scalar integral is either
a $G$ integral, or it can be written as an integral of a certain $G$
after combining all propagators with the same loop momentum
($p$, $q$ or $p+q$) by introducing Feynman parameters.
Where further numerical integrations are needed, the numerical techniques
are similar to those used for the computation of the $g$ and $f$
functions. The main trick is to perform the integration over the remaining
Feynman parameters on a complex path to avoid the eventual singularities
of the integrand. The speed of the integration increases dramatically
if one uses an optimized integration path, along which the integrand is
smooth enough. The path was defined by means of spline functions.

Some scalar integrals need to be evaluated at vanishing external
momentum. In this case ${\cal F}$ vanishes, and the function $g$
can be integrated analytically in terms of Spence functions
\cite{vdBij:2loop:rho,2loop:method}:

\begin{eqnarray}
\lefteqn{g(m_{1},m_{2},m_{3};0) \, =}         \nonumber \\
& & 1 - \frac{1}{2} \log a  \log b  - \frac{a+b-1}{\sqrt{\Delta^\prime}}
     [ \,  Sp(-\frac{u_{2}}{v_{1}}) + Sp(-\frac{v_{2}}{u_{1}})
                                                \nonumber \\
& &   + \frac{1}{4} \log^{2} \frac{u_{2}}{v_{1}}
      + \frac{1}{4} \log^{2} \frac{v_{2}}{u_{1}}
      + \frac{1}{4} \log^{2} \frac{u_{1}}{v_{1}}
      - \frac{1}{4} \log^{2} \frac{u_{2}}{v_{2}}
      + \frac{\pi^{2}}{6}
   \,  ]
      \; \; ,
\end{eqnarray}
where

\begin{eqnarray}
u_{1,2}        & = & \frac{1}{2}
              ( 1 + b - a
                    \pm \sqrt{\Delta^{\prime}} )  \nonumber \\
v_{1,2}        & = & \frac{1}{2}
              ( 1 - b + a
                    \pm \sqrt{\Delta^{\prime}} )  \nonumber \\
\Delta^{\prime}  & = & 1 - 2 (a+b) + (a-b)^{2}
      \; \; .
\end{eqnarray}

Only two masses appear in this calculation: $m_H$, the mass of the Higgs
boson, and $\sqrt{\xi} \, m_W$, the mass of the Goldstone modes, with
the Landau gauge limit taken in the end of the calculation. Therefore
one needs some mass expansions of the $g(m_{1},m_{2},m_{3};0)$ function.

The necessary expansions are given in the following, sometimes
with unnecessary precision:

\begin{eqnarray}
j(M,M,m) & = & x (-1+ \frac{\log x}{2}) + x^{2} \frac{-5+3 \log x}{36}
                 + {\cal O}(x^{3})
  \\
j(M,m,m) & = & \frac{\pi^{2}}{6} + x (-2+2 \log x)
                 + x^{2} (-\frac{3}{2}+\frac{\pi^{2}}{3}+
 \nonumber \\ & &
		 +3 \log x + \log^{2} x)
                 + {\cal O}(x^{3})
  \\
j(m,M,m) & = & - \frac{\pi^{2}}{6} - \frac{\log^{2} x}{2}
               + x (2-\frac{\pi^{2}}{3}-2 \log x - \log^{2} x) +
 \nonumber \\ & &
	       + x^{2} (\frac{11}{2}-\pi^{2}-7 \log x - 3 \log^{2} x)
               + {\cal O}(x^{3})
  \\
j(m,M,M) & = & -2+ \log x - \frac{\log^{2} x}{2}
               + x \frac{13-6 \log x}{18}+
 \nonumber \\ & &
	       + x^{2} \frac{26-15 \log x}{300}
               + {\cal O}(x^{3})
\end{eqnarray}
\begin{eqnarray}
j^{[1,0,0]}(M,M,m) & = & \frac{1}{M^{2}}[
              1 + x (\frac{7}{18} - \frac{\log x}{3})
              + {\cal O}(x^{2}) ]
  \\
j^{[1,0,0]}(M,m,m) & = & \frac{1}{M^{2}}[
              - 2 x \log x
              + {\cal O}(x^{2}) ]
  \\
j^{[1,0,0]}(m,M,m) & = & \frac{1}{M^{2}}[
              - \frac{\log x}{x} - 2 \log x
	      - \frac{2}{3} x (\pi^{2}+12 \log x +
 \nonumber \\ & &
	      + 3 \log^{2} x)
              + {\cal O}(x^{2}) ]
  \\
j^{[1,0,0]}(m,M,M) & = & \frac{1}{M^{2}}[
                \frac{1-\log x}{x}
	      + \frac{7-6 \log x}{18} +
 \nonumber \\ & &
	      + x \frac{37-30 \log x}{300}
              + {\cal O}(x^{2}) ]
  \\
j^{[0,1,0]}(M,M,m) & = & \frac{1}{M^{2}}[
              - 1
	      + x \frac{2-3 \log x}{18}
              + {\cal O}(x^{2}) ]
  \\
j^{[0,1,0]}(M,m,M) & = & \frac{1}{M^{2}}[
                \frac{-1+\log x}{2} +
 \nonumber \\ & &
	      + x \frac{-7+6 \log x}{36}
              + {\cal O}(x^{2}) ]
  \\
j^{[0,1,0]}(M,m,m) & = & \frac{1}{M^{2}}[
                 \log x
	      + x (\frac{\pi^{2}}{3}+4\log x +
 \nonumber \\ & &
	      + \log^{2} x)
              + {\cal O}(x^{2}) ]
  \\
j^{[0,1,0]}(m,m,M) & = & \frac{1}{M^{2}}[
              - \frac{\pi^{2}}{3}-2\log x-\log^{2} x
	      + x(4-4\frac{\pi^{2}}{3}-
 \nonumber \\ & &
	      -12\log x-4\log^{2} x)
              + {\cal O}(x^{2}) ]
  \\
j^{[0,1,0]}(m,M,m) & = & \frac{1}{M^{2}}[
                \log x
	      + x(\frac{\pi^{2}}{3}+7\log x +
 \nonumber \\ & &
	      +\log^{2} x)
              + {\cal O}(x^{2}) ]
  \\
j^{[0,1,0]}(m,M,M) & = & \frac{1}{M^{2}}[
                \frac{-1+\log x}{2}
	      + x \frac{-7+6\log x}{36}
              + {\cal O}(x^{2}) ]
\end{eqnarray}

Here, $x = \frac{m^{2}}{M^{2}}$,
and the following notations were introduced:

\begin{displaymath}
j(m_{1},m_{2},m_{3}) = g(m_{1},m_{2}, m_{3};0) - 1      \; \; ,
\end{displaymath}
\begin{displaymath}
j^{[\alpha,\beta,\gamma]}(m_{1},m_{2},m_{3}) =
\frac{\partial^{\alpha}}{\partial m_{1}^{2 \, \alpha}}
\frac{\partial^{\beta}}{\partial m_{2}^{2 \, \beta}}
\frac{\partial^{\gamma}}{\partial m_{3}^{2 \, \gamma}}
j(m_{1},m_{2},m_{3})
      \; \; ,
\end{displaymath}
and similar for the derivatives of $f$.

One also needs the following relations:

\begin{eqnarray}
j^{[0,1,0]}(M,M,M) & = & \frac{4 \, \sqrt{3}}{3} \, M \, Cl(\frac{\pi}{3})
                         - 2 \, j^{[1,0,0]}(M,M,M)
\end{eqnarray}
\begin{eqnarray}
j(M,M,M) & = & - \frac{2}{\sqrt{3}} \, Cl(\frac{\pi}{3})
  \\
f(M,M,M;-M^{2}) & = & - \frac{3}{2}
  \\
f(M,m,m;-M^{2}) & = & - \frac{1}{2} + {\cal O}(x)
  \\
f^{[0,0,0,1]}(M,m,m;-M^{2}) & = & - \frac{1}{2} \, \frac{1}{M^{2}}
                                  + {\cal O}(x)
      \; \; .
\end{eqnarray}

%############################################################################
%############################################################################
%############################################################################

\newpage

%############################################################################

{\bf Figure captions }

\vspace{2cm}

{\em Fig.1}    The topologies of the one--loop selfenergy diagrams.

\vspace{.5cm}

{\em Fig.2}    The topologies of the two--loop selfenergy diagrams.

\vspace{.5cm}

{\em Fig.3}    The topologies of the one--loop proper vertex diagrams.

\vspace{.5cm}

{\em Fig.4}    The topologies of the two--loop proper vertex diagrams.

\vspace{.5cm}

{\em Fig.5}    Triangular diagrams which display endpoint singularities.
               The solid line denotes the Higgs, and the dashed
	       one the Goldstone bosons.

\vspace{.5cm}

{\em Fig.6}    The radiative corrections to the partial decay width
               of the Higgs to vector bosons in the one--loop (solid line)
	       and the two--loop (dashed line) approximations as
	       a function of the mass of the Higgs boson.

\end{document}